\documentclass[aps,pra,twocolumn,showpacs,preprintnumbers,amsmath,amssymb]{revtex4-1}

\usepackage{graphicx}
\usepackage{dcolumn}
\usepackage{bm}
\usepackage{natbib}

\usepackage{amsmath}
\usepackage{braket}
\usepackage{mathrsfs}
\usepackage{amsfonts}
\usepackage{color}
\usepackage{tabularx}
\usepackage{longtable}

\setcitestyle{square}

\begin{document}

\title{Spin Relaxation in GaAs: Importance of \\Electron-Electron Interactions}

\author{Gionni Marchetti} 
\email{gm597@york.ac.uk}
\author{Matthew Hodgson}
\email{matthew.hodgson@york.ac.uk}
\author{James McHugh}
\email{jgm511@york.ac.uk}
\author{ Roy Chantrell}
\email{roy.chantrell@york.ac.uk }
\author{Irene D'Amico}
\email{irene.damico@york.ac.uk}
 
\affiliation{
Department of Physics, University of York, York, Heslington YO10 5DD, UK\\
}

\begin{abstract}
We study spin relaxation
in n-type bulk GaAs, due to the Dyakonov--Perel mechanism, using
ensemble Monte Carlo methods.
Our results confirm that spin relaxation time increases with the electronic density in the regime of 
moderate electronic concentrations and high temperature.
We show that the electron-electron scattering in the non-degenerate regime significantly
slows down spin relaxation. This result supports predictions by Glazov and Ivchenko.
Most importantly, our findings highlight the importance of many-body interactions for spin dynamics: we show 
that only by properly taking into account electron-electron interactions within the simulations, results for 
the spin relaxation time---with respect to both electron density and temperature---will reach good {\it quantitative} 
agreement with corresponding experimental data. Our calculations contain no \linebreak fitting parameters.
\end{abstract}

\maketitle

\section{Introduction}\label{sec:Introduction}

Recently, spin coherence in semiconductors has been the focus of both theoretically \cite{glazov2002,glazov2003,jiang2009} and
experimentally \cite{oertel2008,romer2010} research. A key aim is to achieve a clear understanding of spin decoherence phenomena. This is very important for the emerging field of spintronics, whose goal is
to exploit the electron spin, in addition to its charge, within electronic devices. Spin-based devices promise new useful applications in electronics and quantum information \cite{bader2010};
therefore, we wish to control, manipulate and detect electronic spins efficiently, provided that their lifetimes are long enough. The long
spin decoherence times measured in semiconductors \cite{kikkawa1998} have made them the subject of intense research. In particular,
the n-type bulk GaAs semiconductor has been shown to be a suitable material for spintronics, as it provides the easy availability of high-quality samples and the possibility of using time-resolved optical techniques for exciting and
detecting spin-polarized electrons \cite{dyakonov2008}.

 In n-type bulk GaAs, the
main sources of spin relaxation are Elliot--Yafet and Dyakonov--Perel (DP) mechanisms \cite{dyakonov2008}, which depend on the spin-orbit interaction. In the Elliot--Yafet mechanism, spin-orbit interaction causes spin depolarization via spin-flips during the carrier scattering events. The spin relaxation due to the DP mechanism follows from the energy splitting, for any non-zero value of the wavevector, of the spin-up and spin-down states. This is
 present in solids that lack bulk inversion symmetry, like GaAs \cite{dresselhaus2008}. This energy splitting gives rise to an effective magnetic field,
whose Larmor frequency depends on the carrier's momentum. Therefore, each electronic spin will precess at a different, momentum-dependent rate.
In the range of the low-to-medium doping concentrations and high temperatures
considered in this paper, DP becomes the dominant mechanism for spin decoherence.

Very recently, the ensemble Monte Carlo (EMC) method has been equipped for dealing with spin transport \cite{saikin2005,sheetal2011,kamra2011}. EMC is
a stochastic method devised to solve numerically the Boltzmann equation for charge transport in semiconductors \cite{jacoboni1989}.
 Here, we improve the treatment within EMC \linebreak of many-body interactions;
 see Sections \ref{sec3} and
\ref{sec:Methods}.
We will use EMC to estimate the effect of electron-electron scattering on the spin relaxation time (SRT) and present results for n-type bulk GaAs at relatively
high temperatures ($280\le T\le 400$~K) and low-to-moderate doping concentrations ($n_i=10^{16}$ to $2.5\times 10^{17}$ cm$^{-3}$).

Our results display good to very good agreement with
the experimental results by Oertel {\em et al.} \cite{oertel2008} and with no adjustable parameters. In particular, we use the same spin-orbit coupling value, $21.9$ eV {\AA}$^{3}$, suggested in the experimental paper. To our knowledge, this is the first time that EMC simulations can quantitatively reproduce spin-relaxation experimental results, and we will discuss the importance, to this aim, of properly taking into account electron-electron interactions.

Our results also confirm that, in the non-degenerate regime, SRT increases with the electron density, both including or excluding electron-electron
scattering \cite{jiang2009}. The latter is due to the fact that by increasing the doping concentration, the electron-impurity scattering rate increases and, consequently, the related motional narrowing effect.

Finally, our findings suggest that the prediction made for two-dimensional systems by Glazov and Ivchenko \cite{glazov2002, glazov2003}, that electron-electron scattering slows down the SRT via motional narrowing, can be extended to the three-dimensional case.

\section{Physical Model}\label{sec:model}

We study carrier and spin dynamics in n-type bulk GaAs considering a single parabolic energy band (the central $\Gamma$ valley), which determines the
effective isotropic electron mass $m^{\ast}_{lab}=0.067 m_{e}$, $m_{e}$ being the bare electron mass. This approximation is justified, as we do not consider highly energetic electrons excited by a strong electric field, so that inter-valley scattering can be discarded. We include only normal-type scattering
events, such as Umklapp processes, that are negligible in direct-gap doped semiconductors. The scattering mechanisms considered are electron-longitudinal acoustic
phonon scattering, electron-longitudinal polar optical phonon (POP) scattering, electron-single charged ionized impurity scattering in the Brooks--Herring
approach \cite{jacoboni1989, chattopadhyay1981} and finally electron-electron scattering. Piezoacoustical interaction is not included, because it becomes relevant
for GaAs samples only at low temperatures \cite{ridley2000}. The scattering rate for the electron-longitudinal acoustic phonon collisions is determined by the
acoustic deformation potential \cite{jacoboni1989} in elastic approximation, as inelastic absorption/emission processes are important only at low temperatures \cite{jacoboni1989}. The electron-longitudinal
POP scattering rate
(Fr\"{o}hlich interaction) \cite{mahan2000} includes absorption and emission processes with a threshold energy of $35$ meV, making it the only dissipative process
in our model. Phonons are considered at equilibrium at the lattice temperature, $T$. Following \cite{pugnet1981,leo1991}, the screening of the electron-phonon interactions are not taken into account
in the present work.

The scattering rates of electrons from an initial state, $\Ket{i}$, to a final state, $\Ket{f}$, are calculated to first order, according to the Fermi golden rule:

\begin{equation}
\label{eq:FGR}
 R_{i\rightarrow f}= \frac{2\pi}{\hbar} |\Bra{f}  V \Ket{i}|^{2} \delta (E_{f}-E_{i}) \,
\end{equation}
where $V$ is the scattering potential (considered as a perturbation) and $E_{f}$ and $ E_{i}$ are the final and initial energy, respectively.

By neglecting the exchange and correlation effects, Coulomb interaction between two charges in a homogeneous
electron gas is usually estimated using the random phase approximation (RPA) \cite{giuliani2005}, giving rise to
an effective screened potential, $V_{sc}$, whose Fourier components are:

\begin{equation}
 V_{sc}(q,\omega)=\frac{v_{q}(q)}{\epsilon(q,\omega,T)} \,
\end{equation}

Here, $\epsilon(q,\omega,T)$ is the temperature-dependent dielectric function, $v_{q}=e^{2}/\varepsilon q^{2}$  the Fourier components with wavevector $q$ of the bare Coulomb potential and $\varepsilon$ is the material dielectric constant, $\varepsilon=12.9 \varepsilon_{0}$, for GaAs. We approximate $\epsilon(q,\omega,T)$ with the long-wavelength limit of its static counterpart at finite temperature $\epsilon(q=0,\omega=0,T)$. This
 is equivalent to the long-wavelength limit of the linearized Thomas--Fermi approximation (LFTA)
$\epsilon(q,\omega=0,T)= 1+(\beta_{TF}^{2}/q^2)$.
 We use Dingle's finite temperature LFTA for n-type semiconductors \cite{chattopadhyay1981, dingle1955}, which determines the inverse screening length, $\beta_{TF}$, from the following equation:

\begin{equation}
\label{eq:screening}
\beta_{TF}^{2}=\frac{n_{e} e^{2}}{\varepsilon k_{B}T}\frac{\mathscr{F}_{-1/2}(\mu/k_{B}T)}{\mathscr{F}_{1/2}(\mu/k_{B}T)} \,
\end{equation}

\begin{figure}

\begin{center}
\includegraphics*[scale=0.5]{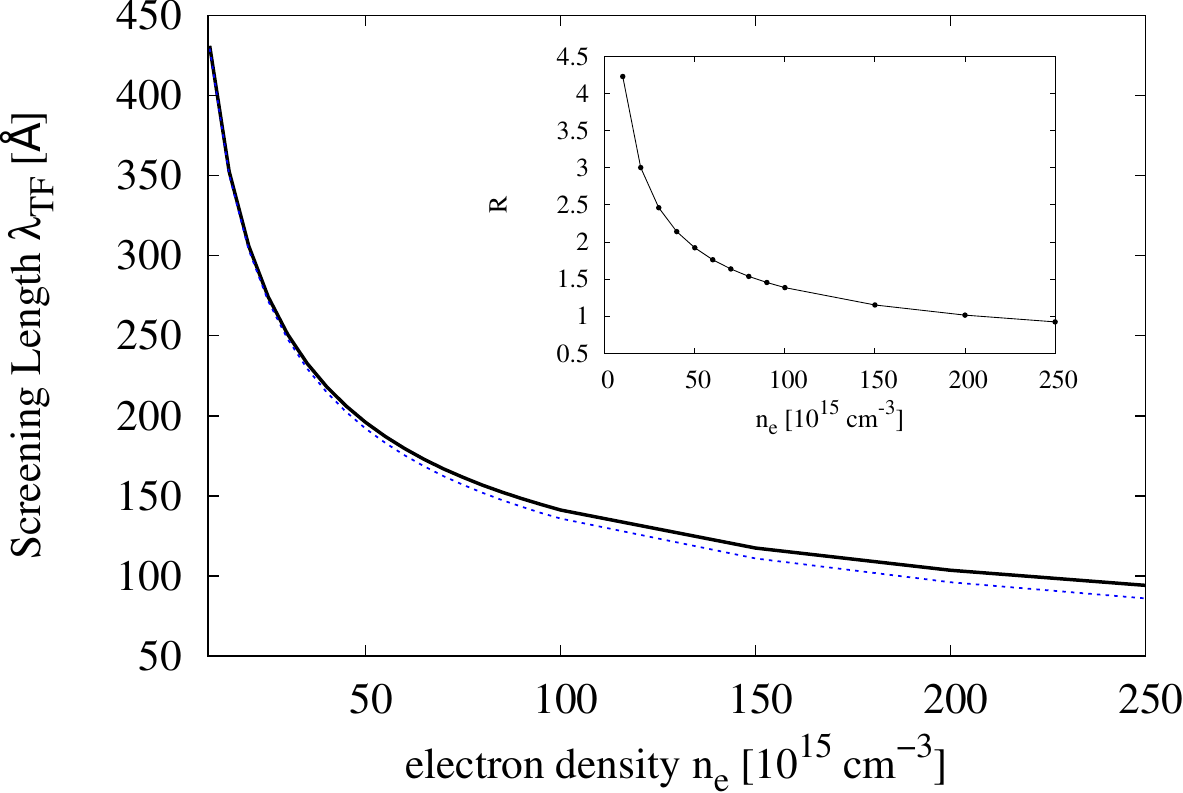}
\end{center}

\caption[] {(\textbf{Main panel}) Screening length (solid line) and Debye--H\"{u}ckel screening length (dashed line) \emph{versus} electron density for an n-type GaAs at $T=300$ K. Here, $n_{e}=n_{i}$, the
latter being the impurity concentration. (\textbf{Inset}) $R$ from Equation (\ref{eq:born1}) \textit{versus} electron density. The parameters are as in the main panel.\vspace{9pt} }\label{fig:screening}
\end{figure}

Here, $n_{e}$ is the electronic concentration, $e$ the elementary charge,
$k_{B}$ the Boltzmann constant, $\mu$ is the chemical potential of the electronic ensemble and $\mathscr{F}_{j}$ denotes the Fermi--Dirac integral of order $j$

\begin{equation}
\mathscr{F}_{j}(x)= \frac{1}{\Gamma(j+1)}\int_{0}^{\infty} \frac{t^{j}}{e^{t-x}+1}dt \,
\end{equation}
with $ x \in \mathbb{R} $ and $\Gamma$ Euler's Gamma function.
In the non-degenerate regime, Equation (\ref{eq:screening}) reduces to the usual Debye--H\"{u}ckel inverse screening length. In Figure \ref{fig:screening}, we plot the values
of the screening length $\lambda_{TF}=1/\beta_{TF}$ calculated according to Equation (\ref{eq:screening}) against the electron density.

\begin{figure}

\begin{center}
\includegraphics*[scale=0.6]{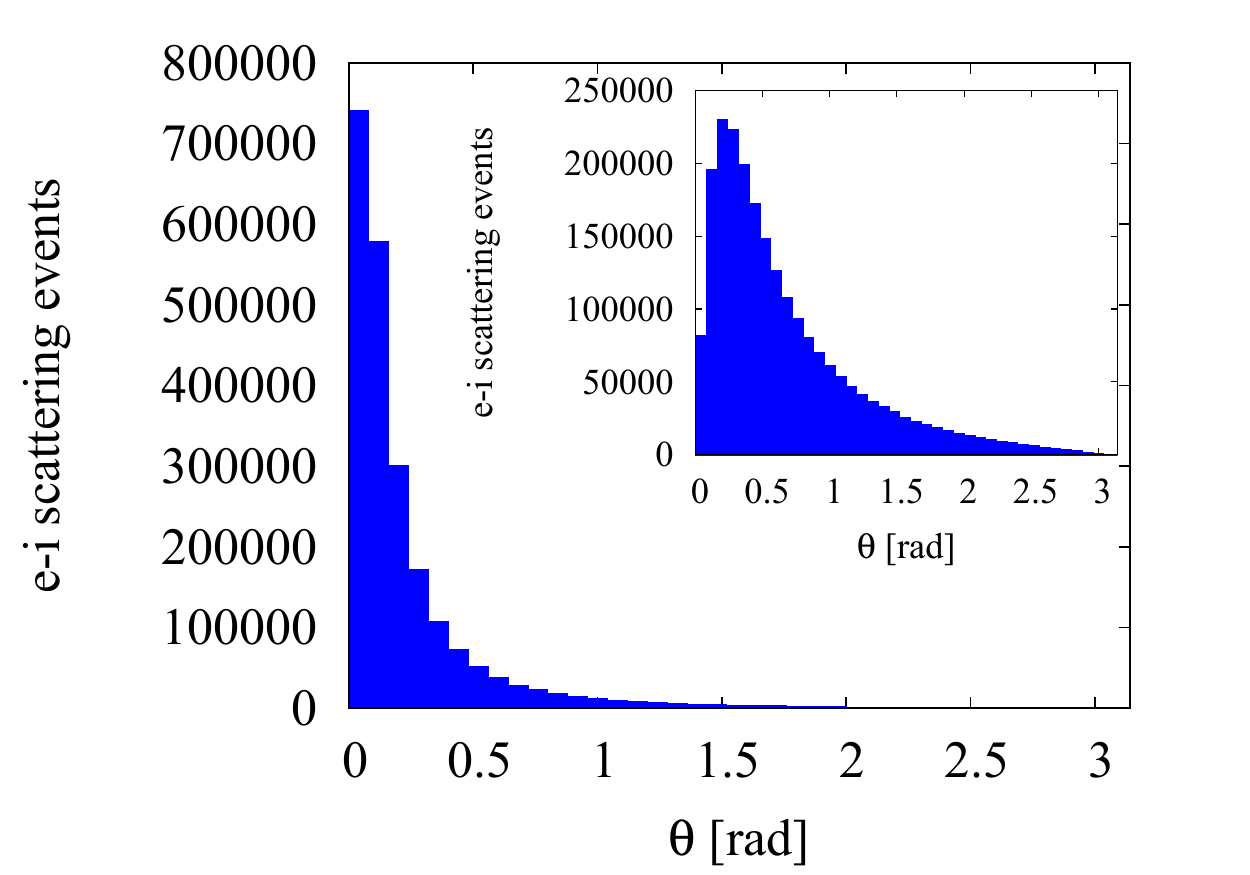}
\end{center}

\caption[] {Histogram of the number of the electron-impurity scattering events
against the scattering angle, $\theta$. The simulation includes $5000$ particles at $T=300$ K. Electron-electron scattering is included in the simulation.
The electron densities are $n_{e}=10^{16}$ cm$^{-3}$ (\textbf{main panel}), and $n_{e}=2.5\times 10^{17}$ cm$^{-3}$ (\textbf{inset}). Note the different scales on the $y$-axis.\vspace{-6pt}
\label{fig:angular_distribution}}
\end{figure}

For Equation (\ref{eq:screening}) to hold, the momentum, $q$, transferred between colliding electron and
impurity must remain small \cite{sanborn1995}. As the electron-impurity process is treated as elastic,
$q$ is given by:

\begin{equation}
q=2 v \sin \left(\theta/2 \right) \,
\end{equation}
where $v$ is the magnitude of the electron (group) velocity and $\theta$ is the scattering angle. Insofar as the electron-impurity scattering favours
small scattering angles, $q$ remains small, and therefore, Equation~(\ref{eq:screening}) gives an accurate approximation of the dielectric function in RPA.
The electron-impurity scattering angular distribution from our simulations confirms that the LTFA is a good approximation in the regime under investigation, especially at low densities,
as the majority of the scattering events happens at small angles; see Figure \ref{fig:angular_distribution}.

\section{Screened Electron-Electron Interaction}\label{sec3}

Within the RPA, Bohm and Pines \cite{pines1952} have shown that it is possible to split the Coulomb interaction between electrons
in two contributions: the first from the collective long-range behaviour (the electron-plasmon interaction) and the second
equivalent to a screened Coulomb interaction between individual electrons. In the present work, we consider only the latter, as the electron-plasmon
scattering becomes important in GaAs for higher electronic concentrations than considered here \cite{jacoboni1989}. Using LTFA, the
electron-electron interaction may then be approximated by the following screened (Yukawa-type) Coulomb potential:

\begin{equation}
\label{eq:ee-potential}
 \emph{v}_{12}(|\mathbf{r}_{1}-\mathbf{r}_{2}|)=\frac{e^{2}}{4 \pi \varepsilon |\mathbf{r}_{1}-\mathbf{r}_{2}|}
 e^{-\beta_{TF} |\mathbf{r}_{1}-\mathbf{r}_{2}| } \,
\end{equation}
where $\mathbf{r}_{1}, \mathbf{r}_{2}$ are the spatial coordinates of the colliding electrons. Only binary electron-electron collisions are
considered here, as they are the most likely and effective scattering events. The quantum states of mobile electrons should be localized
wave packets, but from the perspective of scattering \linebreak theory \cite{joachain1987}, the results are equivalent to those obtained using plane waves. Using this property, electron-electron scattering rates in the non-degenerate regime could be calculated using \cite{jacoboni1989}:

\begin{equation}
 \label{eq:ee-scattering}
 w_{ee}(\mathbf{k_0})=\frac{m^{\ast}e^{4}}{\hbar^{3}V_{cr} \varepsilon^{2}}\sum_{\mathbf{k}} f_{\mathbf{k}}
 \frac{|\mathbf{k}-\mathbf{k_0}|}{\beta_{TF}^{2}\left[ |\mathbf{k}-\mathbf{k_0}|^{2}+\beta_{TF}^{2} \right]^{2}} \,
\end{equation}
where $V_{cr}$ is the volume of the crystal, $f_{\mathbf{k}}$ is the electron occupation probability (or distribution function), in general, unknown, $\mathbf{k_0}$ is
the wave-vector of the colliding electron and the sum runs over all the other electrons in the ensemble.

Within the EMC method, for any given scattering event, once the electron partner of wavevector $\mathbf{k}$,
involved in the collision is chosen, the final states, $\mathbf{k'_0}$, $\mathbf{k'}$, of the colliding electrons can be determined from the conservation
of total energy and momentum and from the scattering angular distribution, \linebreak$P(\theta)$ \cite{jacoboni1989}:

\begin{equation}
 P(\theta)d\theta= C \frac{\sin \theta d\theta}{\left[ g^{2}\sin^{2}\left(\theta/2 \right)+\beta_{TF}^{2} \right]^{2}} \, \label{eq:ang_distr}
\end{equation}

Here, $g$ denotes the magnitude of the vector $\mathbf{g}=\mathbf{k}-\mathbf{k_{0}}$, $\theta$ is the angle between $\mathbf{g}$ and its final state
$\mathbf{g'}=\mathbf{k'}-\mathbf{k'_{0}}$ and $C$ is a normalization constant:

\begin{equation}
 C=\frac{\beta_{TF}^{2} \left(g^{2}+\beta_{TF}^{2}\right)}{2} \,
\end{equation}

\vspace{3pt}The expression for the scattering rate in Equation (\ref{eq:ee-scattering}) arises from our ignorance about the scattering partner in electron-electron collisional events. This
explains the presence of the distribution function in Equation (\ref{eq:ee-scattering}). However, in our simulations, after having determined the scattering type, we explicitly determine the electron partner
from the ensemble. We do so choosing the second electron via a flat distribution within a sphere of radius $\lambda_{TF}$ centred on the colliding electron; see Section
\ref{sec:Methods}. This procedure removes our ignorance about the scattering partner involved in the
collisional event and, at the same time, allows us to retain the Bohm and Pines physical picture of individual particles involved
in collisions. Then, it follows that we can compute the e-e (electron-electron) scattering rate in a simpler way,
using two other ingredients: the Born Approximation and the non-degenerate nature of the system at hand.

First of all, we note that the Fermi golden rule entails first order Born approximation (BA) (usually simply referred to as \textquotedblleft Born
approximation \textquotedblright). From Equation (\ref{eq:FGR}), we can conclude that the scattering rate in BA must not be sensitive to the sign of the potential, \emph{i.e.}, there is no difference between an attractive and repulsive potential, as long as the magnitude of the charges involved is the same.
Secondly, if the antisymmetry of the colliding electrons (non-degenerate regime) and the internal structure of the \linebreak single-ionized impurities may be ignored,
 we may use the electron-impurity (e-i) scattering rate in the Brooks--Herring approach also for the case of the e-e scattering rate. What now
 differentiates between e-i and e-e scattering rates are the different effective masses involved in the collision and, consequently,
the different reduced masses and energies associated with the particles' relative motion.

All of this considered, assuming a parabolic band and a Yukawa screened potential, the formula for e-e scattering we use in our simulation is the following \cite{jacoboni1989}:
\begin{equation}
w_{ee}(E)=\frac{2^{\frac{5}{2}}\pi n_{e} e^{4}}{(4 \pi \varepsilon)^{2} \sqrt{m^\ast}E_{\beta}^{2}}\frac{\sqrt{E}}{1+4 E/E_{\beta_{TF}}} \, \label{w_ee_JL}
\end{equation}
where $E=E_{lab}/2$ and $m^{\ast}=m^{\ast}_{lab}/2$ are the energy and the effective mass of the colliding electron associated with the relative motion, $E_{lab}$ is the energy in the laboratory frame and $E_{\beta_{TF}}$ is
defined by:

\begin{equation}
 E_\beta=\frac{\hbar^{2}\beta_{TF}^{2}}{2 m^\ast} \,
\end{equation}

\vspace{3pt}In the following, we shall use $w_{ee}(k)$ or $w_{ee}(E)$ or $w_{ee}(v)$ when referring to e-e scattering rates written in
terms of wavevector, energy or velocity variables, respectively. The wavevectors, energies and velocities of the electrons involved in e-e
scattering should be thought of as wavevectors, energies and velocities associated with the relative motion.

\vspace{-3pt}\section{The Born Approximation}
\label{sec:BA}


There are some important consequences about using the BA that we wish to recall.

The BA is well satisfied for sufficiently
fast carriers assuming a weak scattering potential. It is indeed a high-energy approximation.
At low energy ($k a_{0}\ll 1$, where $k$ is the magnitude of the colliding electron wavevector and $a_{0}$ is the range of the scattering potential),
a sufficient condition for the validity of the BA for a central potential (square well) is \cite{mandl1992}:

\begin{equation}
\label{eq:born}
 \frac{m^{\ast}|V_{0}|a_{0}^{2}}{\hbar^{2}} \ll 1 \,
\end{equation}
where $V_{0}$ is the typical strength of a short-range central scattering potential, $V$. For an attractive potential, the inequality Equation (\ref{eq:born}) means that the potential, $V$, is not strong enough to form bound states.

In the case of electron-electron scattering,
assuming a screened Coulomb potential, the  Equation (\ref{eq:born}) becomes \cite{schiff1968}:

\begin{equation}
\label{eq:born1}
 R=\frac{m^{\ast}_{lab} e^{2} \lambda_{TF}}{4 \pi \varepsilon \hbar^{2}}=\frac{ \lambda_{TF}}{a^\ast_B} \ll 1 \,
\end{equation}
with $a^\ast_B=(4\pi\hbar^{2} \varepsilon )/(e^{2}m^{\ast}_{lab})$ the effective Bohr radius.

The inequality in Equation (\ref{eq:born1}) is not satisfied for the range of densities considered here, where $R\stackrel{\sim}{>}~1$ (see Figure \ref{fig:screening}). Here, the Born series, which solves the \linebreak Lippmann--Schwinger equation by iteration,
may need more terms to converge, and for that reason, BA might give values for the differential cross-section that are not entirely reliable. Indeed, a
comparison of differential cross-sections obtained from a Yukawa potential with BA and with exact results obtained by the partial wave method shows that they may be \linebreak significantly different, depending
on the energy, scattering angle, screening and strength of the \linebreak potential \cite{joachain1987}. Unfortunately, the same study shows that results are not improved by including the second term of the Born series for a Yukawa potential, as the differential cross-sections worsen \cite{joachain1987}.

\begin{figure}

\begin{center}
\includegraphics*[scale=0.7]{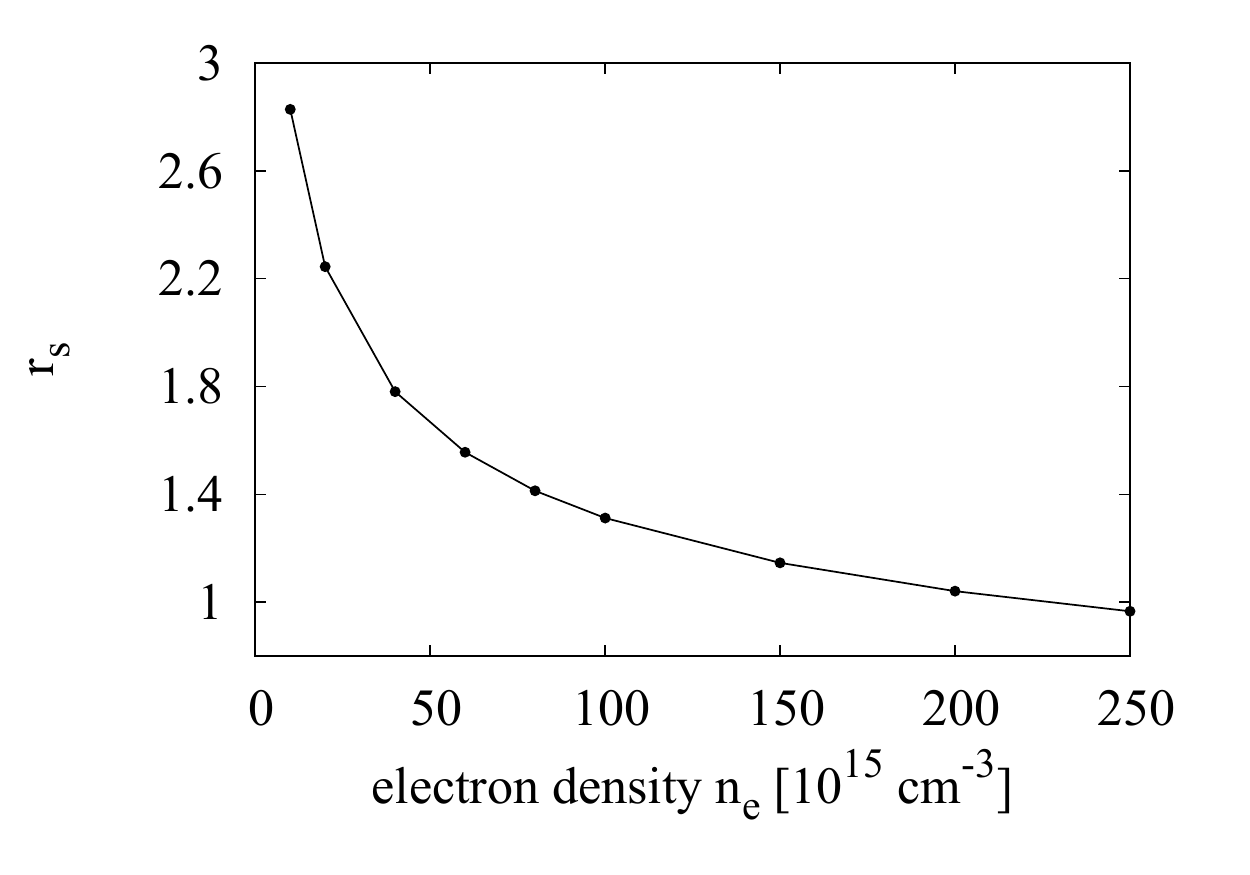}
\end{center}

\caption[] {The behaviour of $ r_s$ \textit{versus} the electron density for an n-GaAs.
 The range of RPA validity is given for values of $r_s$ lesser than one.\vspace{-6pt}}
 \label{fig:rpa}
\end{figure}

There are indications that, when BA is not valid, it tends to overestimate the electron-electron total cross-section and, hence, the e-e scattering. Kukkonen and Smith \cite{kukkonen1973}, using the method of phase shifts, have
found that the electron-electron total cross-section in a metal, like Na (whose average \linebreak inter-electronic distance, $r_s$, is $3.96$ in Bohr radius units),
is overestimated by a factor of two, when assuming a scattering potential, like Equation (\ref{eq:ee-potential}), and including the antisymmetry of the wavefunction of the colliding carriers. This improves over previous results, which did not include the antisymmetry and gave an overestimation of a
factor of five \cite{kukkonen1973}.
The system we are considering is at high temperatures and in a non-degenerate regime; so, the antisymmetry of the wavefunction may be neglected. However, the value of its electron gas parameter, $r_s$, in effective Bohr radius units is similar, with $r_s\stackrel{>}{\sim}1$; see Figure~\ref{fig:rpa}. We might then expect BA to overestimate e-e scattering also in our case.
Clearly, how much the scattering is overestimated is a complicated issue, which strongly relies on the knowledge of the true interelectronic potential.

\section{Ensemble Monte Carlo Method}\label{sec:Methods}

To model electronic and spin dynamics in GaAs and to estimate the spin relaxation time, we employ the ensemble Monte Carlo method 
\cite{jiang2009,sheetal2011,kamra2011}.
This is a semiclassical method in that the simulation has both classical and quantum features.
Such a semiclassical approach is valid in the case that the built-in and applied electromagnetic fields are
spatially slowly varying.

EMC consists of particles' classical ``free-flights'', during which the particles may be accelerated by classical forces,
interrupted by scattering events, which alter the particles' momentum. The probability of such scatterings and the momentum
for each particle following a collision is determined computationally using stochastically generated random numbers.

Among the scattering mechanisms we consider (see Section \ref{sec:model}), the scattering of carriers with the longitudinal polar optical phonons is the only source of
 thermal contact with the lattice. For convenience, we also introduce a fictitious scattering, known in the literature as ``self-scattering'', which does not affect
the particle, but simply ensures that the total scattering rate remains constant \cite{jacoboni1989}.

The free flight time, $\tau$, for a particle is calculated as:\vspace{-6pt}
\begin{equation}
\tau = -ln(r_1)/\Gamma \,
\end{equation}
with $r_1$ a random number generated from a flat distribution between $(0,1)$ and $\Gamma = \Sigma_i w_i^{max}$ the total scattering rate, a constant. Here, $i$ enumerates possible scattering types, and $w_i^{max}$ is the maximum scattering rate possible for
 process $i$. A particle undergoes classical motion for a time, $\tau$; upon free
flight termination, a scattering process is identified for that particle by generating another random number $0 \leq r_2 \leq \Gamma$,
and the scattering mechanism, $i$, is chosen when $w_i^{max} \leq r_2 < w_{i+1}^{max}$. We then calculate $w_i(E)$, the scattering rate associated with the energy, $E$, of the incoming particle.
Given another random number $r_3$, if $r_3 \le w_i(E)$, the particle indeed undergoes a scattering of type $i$, and the scattered particle is then assigned an outgoing momentum,
according to the conservation of energy and momentum for the selected scattering process; else, a ``self-scattering'' is assumed, and no momentum update is \linebreak necessary \cite{jacoboni1989}.

\vspace{-3pt}\subsection{Electron-Electron Scattering}\label{sec:scattering}

The electron-electron scattering has to be handled
somewhat differently, as it involves two particles. Traditionally, a number of approaches have been used, including treating the electron as scattering with a fictitious
partner chosen from a Boltzmann distribution or being
allowed to scatter with an actual simulated particle, whose momentum, though, was not updated. This second particle has been usually chosen irrespectively of its distance from the first particle.

In this work, we improve over previous EMC schemes and allow e-e scattering only between electrons that are
within one screening length of each other. In our scheme, both electrons
scatter, and their momenta are both updated. This approach
prevents the unphysical accumulation of energy or momentum prevalent in other methods, as well as the scattering of electrons at opposite ends of the device.

To implement this, we effectively discretized the space into a
grid of cubes of one screening length side.
We keep track of the number of potential scattering events, which include
each particle scattering off any in the same cube or
in any of the neighboring cubes. Each time an electron-
electron event is required, we choose randomly from each of these
potential pairings and check that they are within one screening
length of each other, and if they are, we carry out the
scattering; if they are not within one screening length of
each other, we choose a different electron as the second particle in the scattering.

\subsection{Thermalization}

In order to start collecting data, we need to wait for the electronic system to thermalize to the chosen lattice temperature.
To do so, we initially allow the system to evolve for an appropriate time (thermalization run), during which the only
source of thermal contact with the environment is provided by polar optical phonons.
We note that the thermalization and the corresponding data collection runs always include the same type of scatterings; in particular, they will both include (or not include) electron-electron interactions.

The initial particle configuration (positions and momenta) for the thermalization run is chosen in the following way. The electron positions are generated
randomly inside the bulk semiconductor using uniform pseudorandom numbers.
Their velocity distribution is generated by choosing the \emph{x}, \emph{y} and \emph{z} components independently from a random Gaussian distribution to
reproduce a Boltzmann distribution with an arbitrary temperature of 130 K, which allows us to check that the system correctly relaxes to the lattice temperature.

In order to ascertain that thermalization is reached,
we have checked when the energy distribution of the carriers becomes a Boltzmann distribution function corresponding to the lattice
temperature. Our simulations show that for the range of parameters of interest in this work and when electron-electron interactions are included,
discarding the first $30$ picoseconds from the simulation is sufficient to ensure thermalization:
in particular, close to room temperature, the thermalization for the runs, including electron-electron interactions, appears to be completed after less than five picoseconds. This confirms the crucial role of electron-electron interactions in the thermalization process \cite{leo1991,collet1986}.
We note that when electron-electron interactions are not included, proper thermalization is not reached, as energy gets hardly redistributed within the electron ensemble.

For the results shown in this work, after the thermalization run, we have reset the electronic spins to be fully polarized along one direction, namely,
the $z$-axis, and then started data collections.

\subsection{Spin Dephasing: The Dresselhaus Term}

In bulk n-GaAs at room temperature and for the range of doping densities here considered, the main source of spin relaxation is the Dyakonov--Perel mechanism, a type of spin-orbit interaction. It is due to bulk inversion asymmetry, and it acts as an effective, momentum-dependent magnetic field,
via the so-called Dresselhaus Hamiltonian \cite{fabian2007,dresselhaus2008}:
\begin{equation}
\label{eq:dresselhaus}
H_{D}=\hbar \mathbf{\Omega}(\mathbf{k})\cdot \vec{\mathbf{\sigma}} \,
\end{equation}
where $\vec{\mathbf{\sigma}}$ are the Pauli matrices and the Larmor precession frequency vector, $\mathbf{\Omega}(\mathbf{k})$, is:
\begin{equation}
\label{eq:dresselhaus1}
 \mathbf{\Omega}(\mathbf{k})=\frac{\gamma_{so}}{\hbar}[k_{x}(k_{y}^2-k_{z}^2), k_{y}(k_{z}^2-k_{x}^2), k_{z}(k_{x}^2-k_{y}^2)] \,
\end{equation}

Here, $k_{i}$ are the wavevector components along the cubic crystal axes, $i=x,y,z$, and $\gamma_{so}$ is known as the Dresselhaus coefficient,
whose values are determined using different methods. In GaAs, $\gamma_{so}$ values have been suggested that range from $8.5$ to $34.5$ eV   {\AA}$^{3}$ \cite{fu2008}.
The Dresselhaus Hamiltonian causes the electron spins to dephase with respect to each other, as each electron spin in the conduction band precesses with a different Larmor frequency $ \mathbf{\Omega}(\mathbf{k})$, which depends on the specific electron's momentum.

\subsection{Spin Evolution}

In the following, we neglect dipole-dipole interaction between spins.
In this way, during free-flight, the spin of each electron undergoes an individual coherent evolution according to the Schr\"{o}dinger
equation.

Initially, each electron spin is assumed to be polarized in the $z$ direction, after which, the spin relaxes via
the Dyakonov--Perel mechanism, whereby each spinor wavefunction is acted upon by the time evolution operator generated by
the Dresselhaus Hamiltonian, $H_{D}$, in Equation (\ref{eq:dresselhaus}).

The time-evolution operator, $U$, in spin space for a single particle spinor wavefunction, $\Psi$, over the time step, $\delta t$, is:

\begin{equation}
\label{eq:cn}
U(\delta t) = e^{-i H_{D} \delta t / \hbar} \,
\end{equation}

so that the spinor wavefunction, $\Psi\left(t\right)$, at time $ \delta t$ is related to its value at the initial time, $t=0$, by:
\begin{equation}
\label{eq:cn1}
 \Psi\left(\delta t \right)= U(\delta t)\Psi\left(0\right) \,
\end{equation}

In order to integrate numerically Equation (\ref{eq:cn1}), we resort to the Crank--Nicolson (C-N) method \cite{crank1947}. This numerical method integrates
by interpolating between two consecutive time steps; hence:

\begin{equation}
\Psi^{n+1} = \Psi^n - \frac{i \delta t}{2 \hbar} H_{D} (\Psi^n + \Psi^{n+1}) \,
\end{equation}
where $\Psi^n=\Psi\left(n\delta t \right)$ denotes the spinor wavefunction at the n
-th-time step. Then, the C-N method leads to the solution:
\begin{equation}
\Psi^{n+1} = \left( 1 + \frac{i \delta t}{2 \hbar} H_{D} \right) ^{-1} \left(1 - \frac{i \delta t}{2 \hbar} H_{D} \right) \Psi^n
\end{equation}
which is correct up to $\mathcal{O}(\delta t)^4$.

This method is particularly convenient, as the inverse of the spin Hamiltonian can be written analytically, allowing for a significant improvement in computational
efficiency compared to the exact solution, with insignificant loss of precision. The C-N method is particularly good for the problem of spin evolution, as it gives a
unitary evolution of the spinor wavefunction in time; hence, it conserves its norm.
 In contrast to the commonly used Heun scheme, the C-N method has the advantage that we do not need
to renormalise the spinor wavefunction after each time step. The explicit numerical scheme is:

\begin{equation}
\Psi(t=\delta t)= C\left(\begin{array}{cc}
1 - \frac{h^2\delta t^2}{4}-i h_z & ih_x +h_y \\
ih_x-h_y & 1 - \frac{h^2\delta t^2}{4}+i h_z \\
\end{array}\right) \Psi(t=0) \,
\end{equation}
where:

\begin{equation}
 C=1+h^2\frac{\delta t^2}{4} \,
\end{equation}
$h_i$ are the $i$-components of the effective field, given by the Hamiltonian in Equation (\ref{eq:dresselhaus}),

\begin{equation}
 h_{i}=2 \Omega_{i} \,
\end{equation}
and $h^2=\sum_{i=1}^{3} h_{i}^{2}$.

At any given time, we can extract the expectation values of the $S_x$, $S_y$ and $S_z$ components of the individual electron spin operator, $S$, to get the
probability that the spin is aligned along each direction. Finally, this can be averaged over all spins to give the net spin in any direction. As in this work, we are starting from an electronic ensemble fully polarized in the $z$ direction, we will be interested in looking at the time evolution of the $z$-component of the total spin, $S_{z,tot}$. At the $n$-th time step, this is given by:
\begin{eqnarray}
S_{z,tot}\left(n\delta t \right) &=& \frac{1}{N}\sum_{i=1,N} \langle S_z\rangle_i \\
&=& \frac{\hbar}{2N}\sum_{i=1,N} \langle \Psi_i\left(n\delta t \right)|\sigma_z|\Psi_i\left(n\delta t \right)\rangle
\end{eqnarray}
where $N$ is the number of electrons in the simulation and $\sigma_z$ the $z$-Pauli matrix.

\vspace{-3pt}\subsection{Estimating the Spin Relaxation Time}

Using the above methodology, we are capable of simulating the time evolution of the total electronic spin and of its components in the sample.
The quantity of interest to us is
the characteristic spin relaxation time of the material. This can be extracted from the time evolution of $S_{z,tot}$.

We assume that, after a transient period, the spin relaxation behavior in the bulk semiconductor takes \linebreak the form:

\begin{equation}
S_{z,tot} = A e^{-Bt}
\end{equation}

\begin{figure}
\begin{center}
\includegraphics*[scale=0.5]{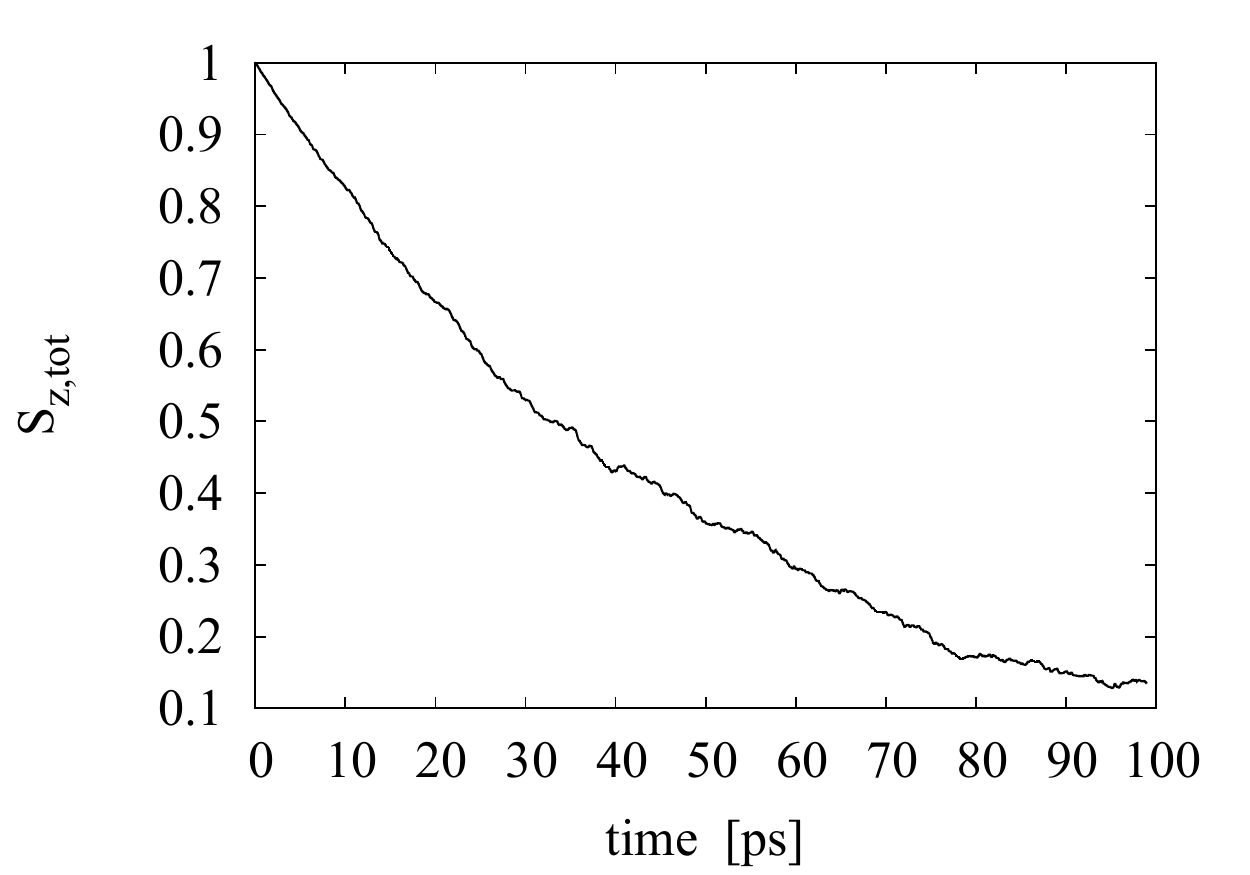}
\end{center}

\caption[] {$S_{z,tot}$ against time for the carrier density $n_{e}=10^{16}$~cm$^{-3}$ at $T$ = 300 K.\vspace{-3pt}}
\label{fig:spin_evolution}
\end{figure}

\vspace{3pt}It is then possible to fit the data from the simulation of the spin time evolution to such a curve (an example is plotted in Figure \ref{fig:spin_evolution})
and produce values for the parameters, $A$ and $B$, in the exponential fit. In particular, the parameter, $B$, has units of $s^{-1}$ and is
identified as the characteristic spin relaxation time of the sample, $B = 1/\tau_s$.
The spin relaxation curve has a behaviour different from an exponential during the first picoseconds; for example, it starts from a
maximum at $t$ = 0, where it then displays a quadratic behavior. We then fit the simulation data exponentially only after this initial transient
time.
From the analysis of the data in the parameter range we are interested in, we see that neglecting the first $10$ ps
of the spin depolarization curve is sufficient for this scope.

\section{Results and Comparison with Experiments}

In this section, we present and discuss our results for the spin relaxation time and compare them to experimental data.

Apart from assuming an exponential decay of the total spin polarization in the $z$-direction, we note that our simulations have no fitting parameters. In particular, the spin orbit coupling value used is not fitted, but we use the value suggested
by Oertel {\em et al.} \cite{oertel2008} for their experimental data: $\gamma_{so}=21.9$ eV {\AA}$^{3}$.

\begin{figure}
\begin{center}
\includegraphics*[scale=0.6]{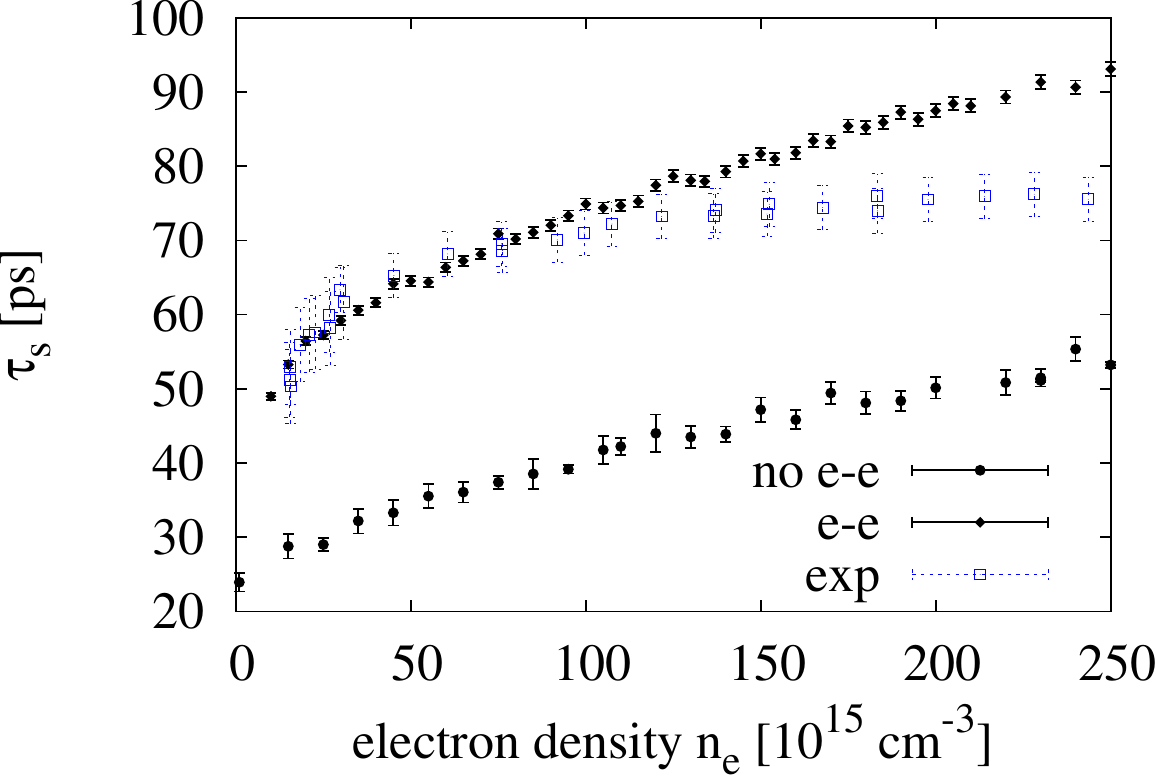}
\end{center}
\caption[] {Results for $\tau_{s}$ {\em vs.} electronic density calculated with and without electron-electron interaction.
Here: $N=25\times 10^3$, $T = 300$ K and $\gamma_{so}=21.9$ eV \AA$^{3}$. The experimental data from
\cite{oertel2008} are plotted, as well (empty square symbols).}
\label{fig:no-ee-and-ee}
\end{figure}

In Figure \ref{fig:no-ee-and-ee}, we plot results from simulations with ($\tau_{s}^{ee}$) and without ($\tau_{s}^{no~ee}$) electron-electron
scattering to examine the effect that the inclusion of electron-electron scattering has on $\tau_s$ at room temperature. In the same figure, we also plot the
experimental results obtained by Oertel {\em et al.} \linebreak in \cite{oertel2008} (empty square symbols).
When we plot $\tau_s$ against the range of densities $n_e$ = 1 $\times$ 10$^{16}$ to \linebreak 2.5 $\times$ 10$^{17}$ cm$^{-3}$, we
see that the inclusion of electron-electron scattering causes a net increases of $\tau_s$ at all densities. Glazov and Ivchenko \cite{glazov2003}
predicted a similar result in the case of a two-dimensional non-degenerate electron gas in GaAs, explaining it with additional motional
narrowing caused by the e-e scattering. Our result suggests that this effect is present also in the three-dimensional case.

Additionally, we notice that the percentage increase of $\tau_s$ with respect to 
its non-interacting approximation decreases with increasing density, from about $\sim$90\% to about $\sim$70\%, remaining though 
always very substantial, even for $n_e=2.5 \times 10^{17}$ cm$^{-3}$. Its absolute increment $\tau_{s}^{ee}- \tau_{s}^{no~ee}$ instead 
increases with the electronic density.

We observe that, when including e-e interaction, our results for densities
$10^{16}$~cm$^{-3} \lesssim n_e\lesssim 10^{17}$~cm$^{-3}$ are in {\it very good agreement with the experimental data} 
reproduced in Figure~\ref{fig:no-ee-and-ee}.

 However, at higher densities, our results for $\tau_{s}^{ee}$ start to overestimate the experimental data for $\tau_s$, reaching a $\sim$20\%
 overestimate when $n_e = 2.5\times 10^{17}$~cm$^{-3}$.

We suggest that the overestimate of $\tau_s$ for $n_e\gtrsim 10^{17}$~cm$^{-3}$ is due to the BA overestimating the e-e scattering rate,
as discussed in Section~\ref{sec:BA}.

We focus now on the effect of temperature on the spin relaxation time.

In order to compare our calculations with other experimental data from \cite{oertel2008}, we consider the temperature r
ange 280~K $\le T \le 400$~K and two (fixed) densities, $n_{e} = 2.7 \times 10^{16}~$cm$^{-3}$ and
$3.8 \times 10^{16}~$cm$^{-3}$.
In both cases, we will consider interacting carriers.

\begin{figure}

\begin{center}
\includegraphics*[scale=0.6]{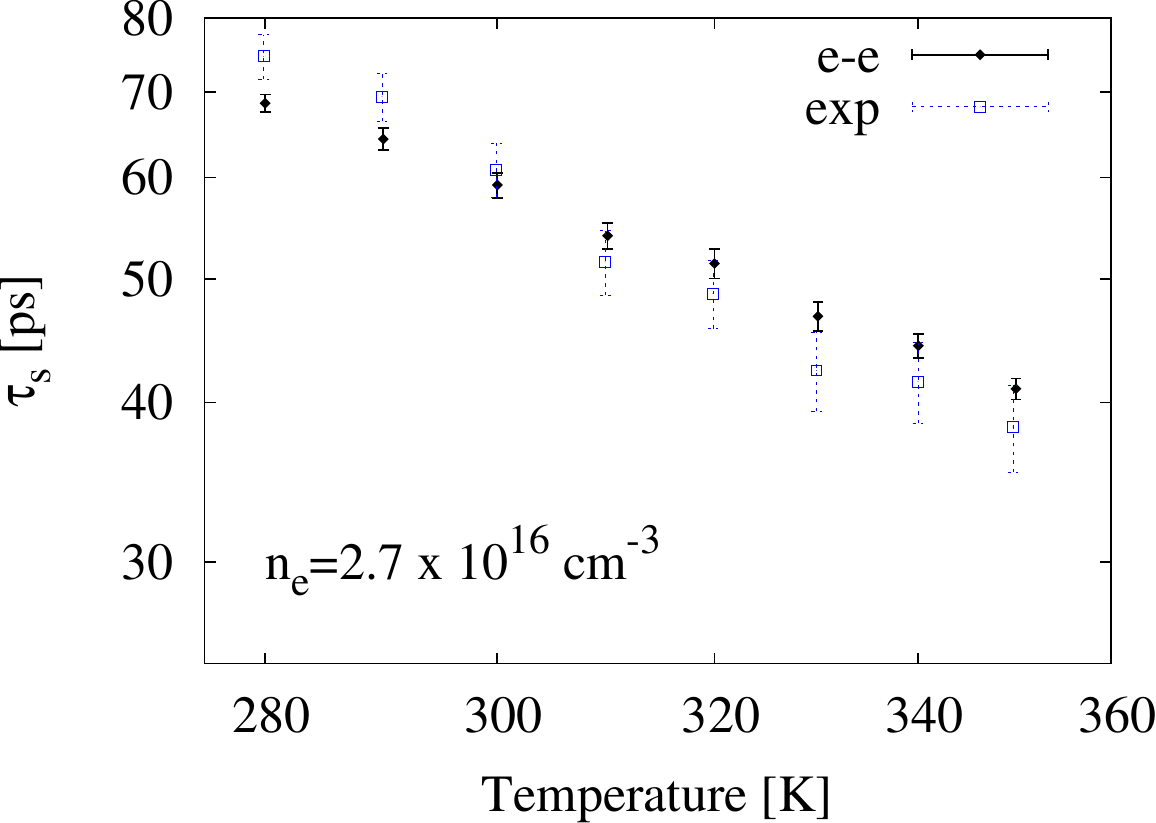}
\end{center}

\caption[] { The spin relaxation time, $\tau_s$, \textit{versus} temperature for the carrier density \linebreak $n_{e}=2.7 \times 10^{16}$ cm$^{-3}$ from simulations, including electron-electron interaction, and from
the experimental results, as obtained in \cite{oertel2008} (empty square symbols). The simulations are done with
$N= 25\times 10^3$ and $\gamma_{so}=21.9$ eV \AA$^{3}$ and include e-e  scattering. Following \cite{oertel2008},
data are plotted on a log-log scale.\vspace{9pt}}
\label{fig:2.7density}
\end{figure}

\begin{figure}

\begin{center}
\includegraphics*[scale=0.6]{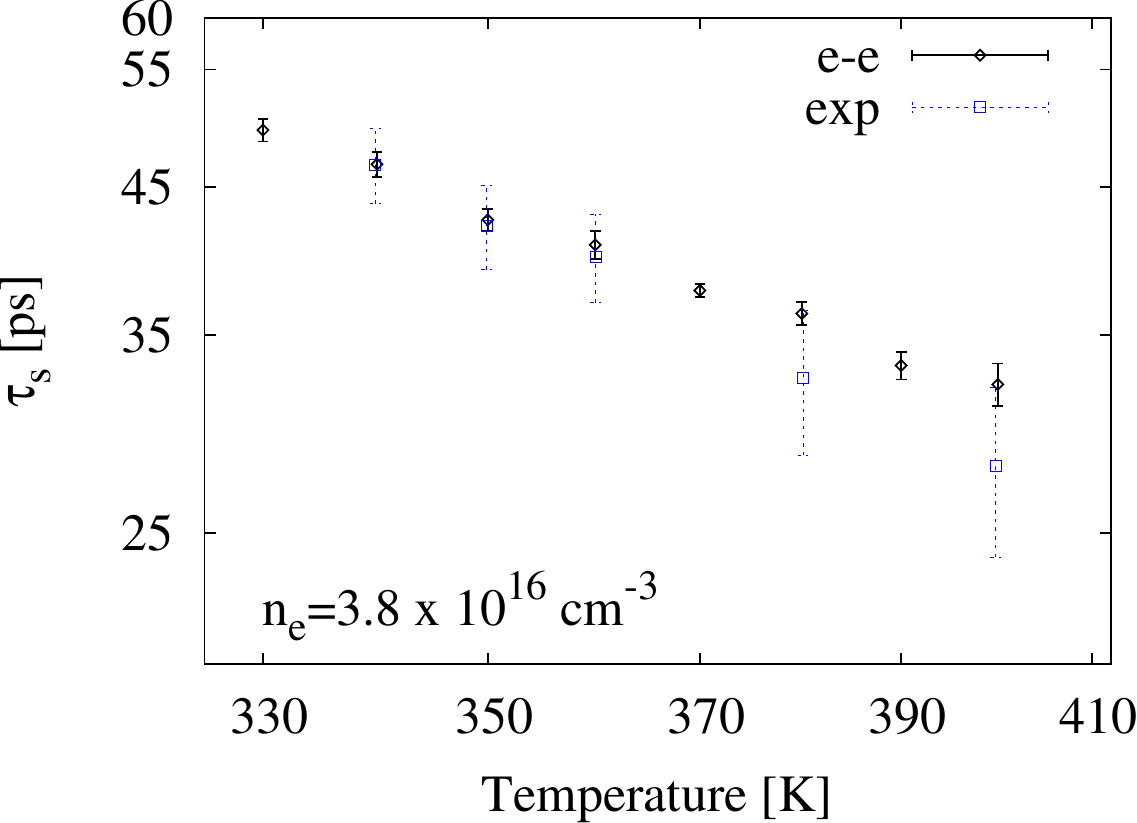}
\end{center}

\caption[] {The spin relaxation time, $\tau_s$, \textit{versus} temperature for the carrier density \linebreak $n_{e}=3.8 \times 10^{16}$ cm$^{-3}$ from simulations, including electron-electron interaction, and from
the experimental results, as obtained in \cite{oertel2008} (empty square symbols). The simulations are done with
$N= 25\times 10^3$ and $\gamma_{so}=$ 21.9 eV \AA$^{3}$ and include e-e scattering. Following \cite{oertel2008}, data are plotted on a log-log scale.\vspace{9pt}
\label{fig:3.8density}}
\end{figure}

In Figures \ref{fig:2.7density} and \ref{fig:3.8density}, we present our results for $n_{e} = 2.7 \times 10^{16}~$cm$^{-3}$ and $n_{e} = 3.8 \times 10^{16}~$cm$^{-3}$ alongside the corresponding experimental data (empty square symbols).
We find good agreement over the entire temperature range between $\tau_s^{ee}$ and
 the experimental data.

\subsection{Dependence on the Value of the Spin-Orbit Coupling}

As noted before, the values of the spin orbit coupling for GaAs found in the literature vary \linebreak greatly \cite{jiang2009}; one of the main points in our work is that we do not treat $\gamma_{so}$ as an adjustable parameter, but simply use the value provided by experimentalists.

In order to let the reader appreciate how valuable this is, and in this respect, how relevant is the good agreement between our data and the experimental ones, in this section, we wish to show how sensible our simulations are with respect to the value of $\gamma_{so}$.

\begin{figure}

\begin{center}
\includegraphics*[scale=0.6]{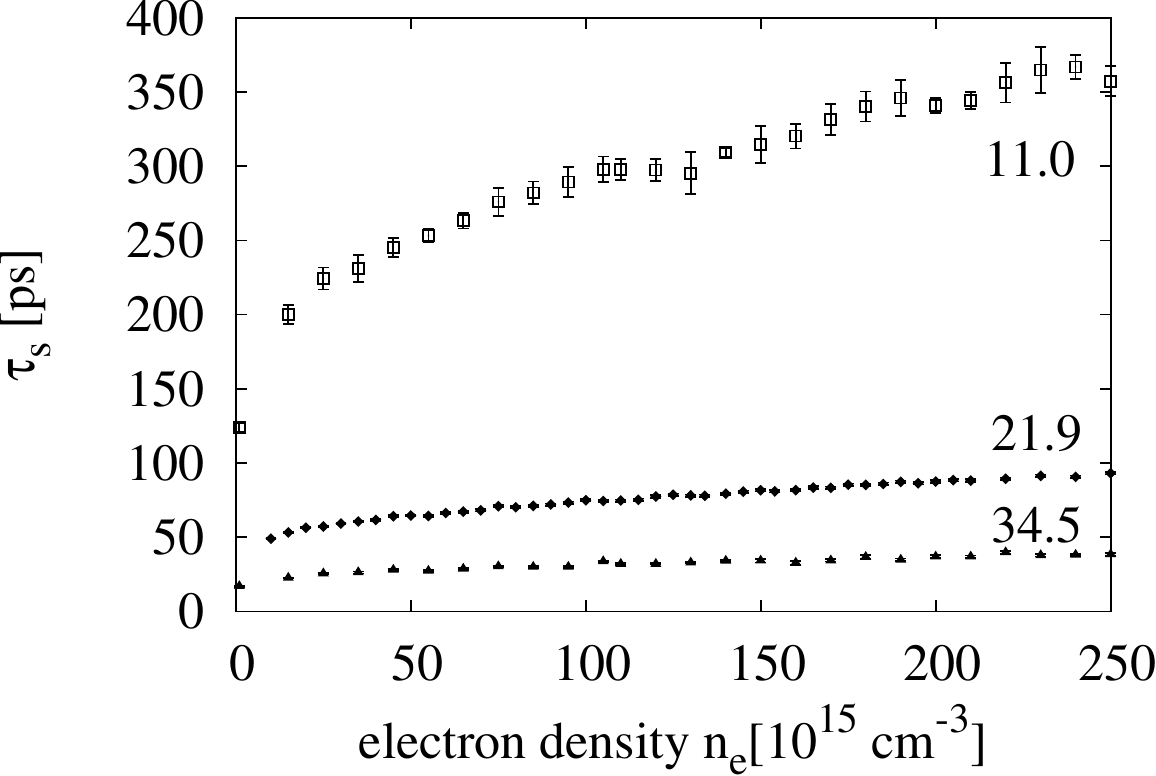}
\end{center}

\caption[] {Spin relaxation time $\tau_{s}$ {\em vs.} density from simulations, including electron-electron interaction, and for three different values of the spin-orbit coupling, $\gamma_{so}=11.0,~21.9,~ \linebreak 34.5$ eV \AA$^{3}$. Other parameters:
$N= 25\times 10^3$ and $T$ = 300 K.\vspace{9pt} }
\label{fig:soc}
\end{figure}

In Figure~\ref{fig:soc}, we plot $\tau_{s}^{ee}$ for three different values of $\gamma_{so}$, all within the range suggested in the literature. It can be seen that by varying $\gamma_{so}$, the results for the spin relaxation time would vary within one order of magnitude, and this for the whole density range here considered.

We think that this is a convincing proof that the very good agreement between our results and the experiments is not accidental, but derives from the improvements we have devised in treating the e-e interaction within the EMC method. These improvements allow us to account properly for the \linebreak electron-electron interaction within the simulations.

\subsection{Statistical Analysis of the Spin Relaxation Time Using Coulomb Differential Cross-Sections}

We wish to understand better the results relative to the e-e curve in Figure~\ref{fig:no-ee-and-ee}. To do so, we focus only on the e-e scattering mechanism,
assuming that the other scattering mechanisms give a correct collisional probability.
 By comparing our calculations with the experimental data, we see that the e-e scattering overestimates $\tau_s$ at higher concentrations. This may be due to the fact that
the e-e scattering itself is overestimated, being that BA is not such a good approximation for low energy carriers; see Section~\ref{sec:BA}. Surprisingly, though, we find very
good agreement with the experimental data for densities lower than $ 10^{17}$~$cm^{-3}$, while, as BA worsens at lower densities, we would expect that the SRT curve we obtained from our calculations lies above the experimental curve for the entire range of densities.

To explain this good agreement in the low density limit, we make some general considerations about Coulomb scattering, RPA and
screening. Going towards low densities, the RPA starts to break
down, which means that in our model, we are no longer allowed to split the e-e interaction into two parts. This can
be also understood by looking at $r_{s}$, as a criterion for the validity of RPA is \cite{bennacer1989}:
\begin{equation}
 r_s \lesssim 1
\end{equation}

From Figure~\ref{fig:rpa}, we see that in our system RPA criterion starts to break down for $n_e\lesssim 1.5\times 10^{17}$~cm$^{-3}$, interestingly a range comparable to the one in which we find agreement between our results for $\tau_s^{ee}$ and the experimental data.
This shows that, at low densities, the potential energy starts to dominate over
the kinetic energy. In other words, the long-range component of the Coulomb interaction becomes relevant, 
and a Yukawa-type potential may be no longer sufficient to realistically describe the \linebreak inter-electronic~potential.

The breakdown of RPA in low electronic densities may affect also the screening length, whose calculation 
strongly relies on this approximation in our model and, consequently, making the scattering probability less reliable.

The RPA breaking down means that e-e scattering should, in the real system, be more effective. However we still use a
Yukawa potential in our calculations; so, as the density decreases, we should be underestimating the e-e scattering and, so, should 
obtain a $\tau_s^{ee}$ smaller than the real $\tau_s$.
However the lower range of density we consider corresponds to the regime where RPA starts to break down (which is compatible with the system $r_s$ values), so that the e-e scattering, which results from our simulations, is accidentally correct. We can think of three regimes.
In the first, with $n_e\gtrsim 1.5\times 10^{17}$~cm$^{-3}$, RPA is appropriate as $r_s\lesssim1$. 
BA works well enough as $R\sim 1$, and as a result, our simulations overestimate the e-e scattering, \emph{i.e.}, $\tau_s^{ee} > \tau_s$.

In the opposite limit ($r_s \gg 1$), RPA is completely inadequate: here, the dominant part of the e-e scattering comes from 
the long range component of the Coulomb interaction, and if a Yukawa potential would still be used, the simulations would
underestimate the e-e scattering; and as a result, $\tau_s^{ee} < \tau_s$. From the trend of $r_s$ (see Figure~\ref{fig:rpa}), this 
should happen for densities $n_e \lesssim 10^{16}~cm^{-3}$, which we do not simulate and which are not realistic, 
because the system becomes an insulator.

The third regime is intermediate and corresponds to
$r_s$ of the order of one, with $r_s > 1$. In this regime, RPA has not completely broken down, but the long-range part of the 
Coulomb interaction starts to become relevant. Using a Yukawa potential then underestimates the e-e interaction, but at 
the same time, the use of BA (which overestimates the e-e interaction) compensates for this; and we get as a result that
$\tau_s^{ee} \sim \tau_s$. By looking at the values of $r_s$ \textit{versus} density (Figure \ref{fig:rpa}), $r_s\gtrsim 1$ for 
the density range $1\times 10^{16}$~cm$^{-3}\lesssim n_e\lesssim1.5\times 10^{17}$~cm$^{-3}$. We indeed 
find that $\tau_s^{ee} \sim \tau_s$ for the density range $1\times 10^{16}~$cm$^{-3}\lesssim n_e\lesssim 1.2\times 10^{17}~$cm$^{-3}$ 
(see Figure \ref{fig:no-ee-and-ee}).

Another way of looking at the previous considerations is that, for low electronic densities, the system differential cross-section, as described by our simulations, is in some way mimicking a bare Coulomb potential one. Because the later is the exact differential cross-section of the system \cite{weinberg2013}, if our simulations are mimicking it, the related scattering probability would not be overestimated and the quantitative agreement with the experimental result explained.

We will now demonstrate that indeed in our simulations and for the low density range, $\sigma^{bare}\approx\sigma^{Y}$, with $\sigma^{bare}$ the bare Coulomb differential cross-section and the $\sigma^{Y}$
Yukawa differential cross-section in BA.
To do so, we will then consider the ratio $\sigma^{ratio}= \sigma^{Y}/\sigma^{bare}$ and determine the conditions, such that $\sigma^{ratio}\approx 1$.

The Yukawa differential cross-section in BA is given by \cite{mandl1992}:

\begin{equation}
\label{eq:yukawa}
 \sigma^{Y}\left(\theta \right)= \frac{ e^{4}}{\left(4\pi \varepsilon\right)^{4}} \frac{1}{\left(E_{\beta_{TF}}+4 E \sin^2(\theta/2)\right)^2}\,
\end{equation}
where $\theta$ is the scattering angle associated with the relative motion.
The bare Coulomb differential cross-section is obtained from Equation (\ref{eq:yukawa}) in the limit of $\beta_{TF}\rightarrow 0$, 
\emph{i.e.}, no screening.
Then, $\sigma^{ratio}$ is:

\begin{eqnarray} \label{eq:sigma_ratio}
 \sigma^{ratio}(\theta, \xi)& =& \frac{\xi^2 \sin^4(\theta/2)}{(1+\xi \sin^2(\theta/2))^2}\,\\
&=& \frac{s^2(\xi;\theta)}{[1+s(\xi;\theta)]^2}\,\label{eq:sigma_ratio2}
\end{eqnarray}
where we have defined the dimensionless quantities $\xi=4 E/E_{\beta_{TF}}= 2 E_{lab}/E_{\beta_{TF}}$
and $s= \xi \sin^2 \left({\theta}/{2}\right)$. From Equation (\ref{eq:sigma_ratio}), we note that $\sigma^{ratio}\le 1$ always.
For a given set of energies, $E$, of the collisional electrons and a threshold value $\sigma^{\ast}$ of $\sigma^{ratio}$ close to unity, there may
exist the set of scattering angles $I_{\theta^{\ast}}=
\{\theta \in \left[\theta^{\ast}, \pi \right]: \sigma^{\ast}\leq \sigma^{ratio} \leq 1 \}$. 
From the angular probability distribution $P(\theta,E)$ (see
Equation~(\ref{eq:ang_distr})), we can determine the probability, $F_{\theta^{\ast}}$, that, for a given $E$, $\sigma^{\ast} \leq \sigma^{ratio} \leq 1$:
\begin{equation}
 F_{\theta^{\ast}}\left(E \right)= \int_{\theta^{\ast}}^{\pi} P(\theta,E) d \,\theta \,
\end{equation}

This integral can be solved analytically, and we get:
\begin{eqnarray}
 F_{\theta^{\ast}}\left(\xi \right)
&=& \frac{\cos^2 \left(\frac{\theta^{\ast}}{2}\right)}{1+\frac{\xi}{4}\sin^2 \left(\frac{\theta^{\ast}}{2}\right)}\\
&=&\frac{4}{4+s(\xi;\theta^{\ast})} \frac{\xi-s(\xi;\theta^{\ast})}{\xi}\,
\end{eqnarray}

Because the system is at equilibrium, we can use the Boltzmann distribution, $f_{B}(E)$, to weight the function, $F_{\theta^{\ast}}$, over
the whole energy spectrum,
giving the probability
 that an e-e collisional event gives $\sigma^{\ast}\le\sigma^{ratio}\le 1$.
By using that for fixed $\sigma^{\ast}$, $s(\xi;\theta^{\ast})$ becomes a constant, $s^*$, which can be determined from Equation~(\ref{eq:sigma_ratio2}); this probability is given by the
integral:
\begin{eqnarray}
 \label{eq:integral}
I &=& 2 \sqrt{\frac{1}{\pi}}\frac{1}{(k_B T)^{3/2}} \int_{E^{\ast}_{lab}}^{\infty} \sqrt{E} e^{-E/k_B T} F_{\theta^{\ast}}(E) dE \, \\
&=&
 \frac{8\alpha^{3/2}}{\sqrt{\pi}}\frac{4}{(4+s^*)} \int_{\frac{2E^{\ast}_{lab}}{E_{\beta_{TF}}}}^{\infty}
 \frac{\xi-s^*}{\sqrt{\xi}}e^{-\alpha \xi} d\xi \,
\end{eqnarray}
where $\alpha=E_{\beta_{TF}}/(2k_B T)$ and the lower integral limit, $E^{\ast}_{lab}$, must be determined. This is the the smallest energy for which it is still possible to obtain $\sigma^{ratio}(\theta, \xi)$ as large as $\sigma^\ast$. In the limit $E_{lab}\to E^{\ast}_{lab}$, we have that $\theta^\ast\to\pi$.
 Imposing the condition $\sigma^{ratio}(\theta^\ast, \xi)\ge\sigma^\ast$, we obtain $s(\xi;\theta^{\ast})\ge \frac{\sigma^{\ast}+\sqrt{\sigma^{\ast}}}{1-\sigma^{\ast}}$ from which, in the limit $\theta^\ast\to\pi$, we get:
\begin{equation}
 \frac{2E^{\ast}_{lab}}{E_{\beta_{TF}}}= \frac{\sigma^{\ast}+\sqrt{\sigma^{\ast}}}{1-\sigma^{\ast}} \,
\end{equation}

\begin{figure}

\begin{center}
\includegraphics*[scale=0.6]{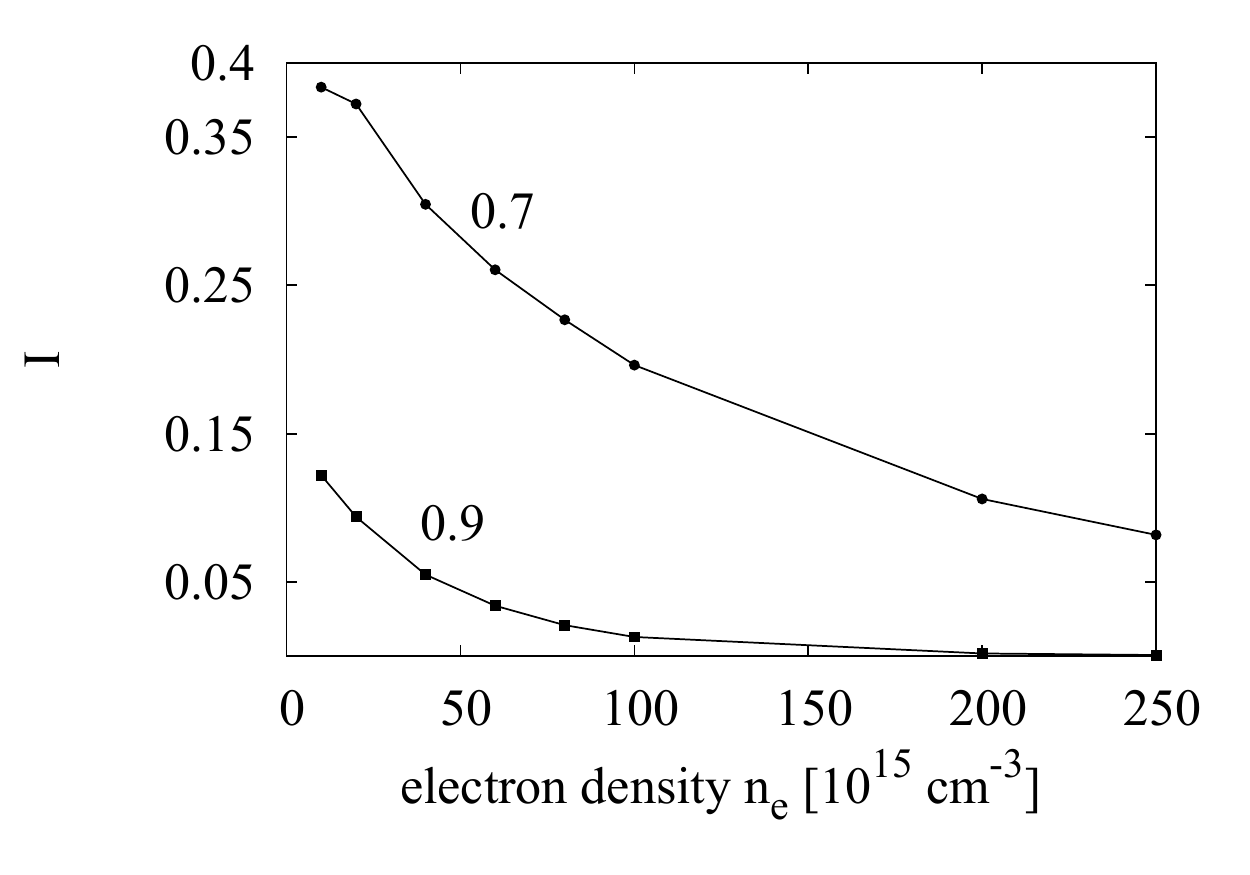}
\end{center}

\caption[] {The integral, $I$, \textit{versus} electron density at $T$ = 300 K for two different values of $\sigma^{\ast}$, $0.7$ and $0.9$, as labelled. \vspace{-6pt}}
\label{fig:sigma}
\end{figure}

\vspace{3pt}The integral, $I$, is a function of the electronic density through $\alpha$, so it is possible to compare the probabilities for
different electronic densities at $T$ = 300 K. We calculate $I$ for $\sigma^{\ast}$ = 0.7 and 0.9; see Figure \ref{fig:sigma}.
The results indicate
that the e-e collisions with a differential cross-section close to the bare one are more favored at lower densities, which proves our point. The curves in Figure~\ref{fig:sigma} show that, for a density as high as $n_e = 2.5\times 10^{17}$~cm$^{-3}$, only 8\% of the total number of carriers would scatter with a differential cross-section, such that $0.7\le\sigma^{ratio}\le 1$.

\vspace{3pt}

\section{Conclusions}

We have improved the treatment of e-e scattering in ensemble Monte Carlo and shown that our method allows one to reproduce, with no fitting parameters,
the experimental results for spin relaxation by \linebreak Oertel {\em et al.} We obtain good agreement over the whole range of electron densities and temperatures considered
experimentally. Our results show that, in order to achieve quantitative agreement with the experiment, it is crucial to properly include e-e interactions
within the simulations. Failure to include many-body interactions leads to greatly underestimating the spin relaxation time.

For the highest electron densities considered, the Born approximation slightly overestimates the e-e scattering rate and, hence,
the corresponding scattering cross-section. This implies a higher probability of having a third electron within the scattering cross-section.
As future work, we wish to study the importance of this spurious ``third-body'' effect on spin dynamics in semiconductors
and evaluate if an appropriate treatment of it can further improve the agreement with the experimental results.

\begin{acknowledgments}

We acknowledge support from EPSRC Grant
No. EP/F016719/1 and from the European \linebreak Community's
Seventh Framework Programme (FP7/2007-2013) under
grant agreement
281043 ``FemtoSpin''. We thank Michael Oestreich for the experimental curves.

Gionni Marchetti and Irene D'Amico analyzed the data and wrote the
manuscript. Matthew Hodgson and Gionni Marchetti together developed
the code for the spin transport. James  McHugh carried out part of the
numerical simulations. Irene D'Amico and Roy Chantrell supervised the
study. All authors contributed to the scientific discussion.

The authors declare no conflict of interest.

\end{acknowledgments}

\bibliography{jap_references}{}
\bibliographystyle{ieeetr}

\end{document}